\documentclass[conference, 10pt]{IEEEtran}

\usepackage{epsfig}
\usepackage{flushend}
\usepackage{fixltx2e}
\usepackage{url}
\usepackage{amsmath}
\usepackage[utf8]{inputenc}
\usepackage{graphicx}
\usepackage{subcaption}

\begin{document}

\title{On GitHub’s Programming Languages}

\author{\IEEEauthorblockN{Amirali Sanatinia, Guevara Noubir}
\IEEEauthorblockA{College of Computer and Information Science\\
Northeastern University, Boston, USA\\
\{amirali,noubir\}@ccs.neu.edu}
}



\maketitle

\begin{abstract}
GitHub is the most widely used social, distributed version control system. It has around 10 million registered users and hosts over 16 million public repositories. Its user base is also very active as GitHub ranks in the top 100 Alexa most popular websites. In this study, we collect GitHub's state in its entirety. Doing so, allows us to study new aspects of the ecosystem. Although GitHub is the home to millions of users and repositories, the analysis of users' activity time-series reveals that only around 10\% of them can be considered active. The collected dataset allows us to investigate the popularity of programming languages and existence of pattens in the relations between users, repositories, and programming languages. 

By, applying a k-means clustering method to the users-repositories commits matrix, we find that two clear clusters of programming languages separate from the remaining. One cluster forms for ``web programming'' languages (Java Script, Ruby, PHP, CSS), and a second for ``system oriented programming'' languages (C, C++, Python). Further classification, allow us to build a phylogenetic tree of the use of programming languages in GitHub. Additionally, we study the main and the auxiliary programming  languages of the top 1000 repositories in more detail. We provide a ranking of these auxiliary programming languages using various metrics, such as percentage of lines of code, and PageRank.
\end{abstract}

\section{Introduction}\label{s:intro}

GitHub is the most widely used social code hosting platform, based on Git, a distributed version control system. It introduces a social aspect to software development where users can browse, fork and even contribute to the projects created and maintained by others. Such platform facilitates agile development and has the potential to address problems such as collaboration, communication, and code conflicts. Furthermore, it provides a social platform similar to Twitter, for users to interact. For example, users can follow another user or mention them in discussions.

Currently, GitHub is the home to more than 10 million registered users and over 16 million public repositories. It has more users and hosts more projects than other source code hosting platforms, such as SourceForge (324000 projects)~\cite{SourceForge}, Google Code (250000 projects)~\cite{GoogleCode}, or Launchpad (32000 projects)~\cite{Launchpad}. The number of users on GitHub has been increasing exponentially until 2014. For instance, the number of users created in 2013 and 2014, is twice more than the number of users created from 2007 to 2012, combined. However, it seems like 2014 is an inflection point in GitHub's growth. GitHub is ranked 98 on Alexa, as of May 2015. The majority of the visitors~\cite{alexa} are from the United States (19.6\%), followed by India (15.5\%), China (8.5\%), Russia (3.3\%) and Brazil (3.1\%). It is a truly geographically diverse collaboration platform where users can contribute to the development of open-source software. Not only, it is a popular choice between developers, scientists and hobbyist, even enterprises such as Lockheed Martin, Microsoft, LivingSocial, VMware, and Walmart also use GitHub~\cite{gitenterprise}.

GitHub is rising as a platform for social open-source software development, and previous studies have looked at its social aspects~\cite{Begel:2013:SNM:2478553.2478856}. However, they are either limited to surveys, or are based on a relatively smaller sample of repositories. A body of literature~\cite{Ray:2014:LSS:2635868.2635922,Rahman:2014:IPR:2597073.2597121} mined software repositories such as GitHub, but they had a smaller scope. To the best of our knowledge we are the first to thoroughly investigate the programming languages relationship in Open Source Software development on GitHub, we are also the first to perform a large scale analysis of GitHub's dynamics, spanning 8 years of contributions. From 2007, when the first repository was created by the first registered user (co-founder, Tom Preston-Werner), until the end of 2014. Such large scale holistic dataset allow us to investigate research problems more conclusively.

In this work, we investigate the state of modern software development. Today's software is not only developed by professional software engineers, but by a large diverse group of users some of whom are considered amateur and hobbyist. Modern software artifacts also use more than one programming language, many of them utilize an array of programming language to achieve different objectives. It is of paramount importance to understand the relationship of programming languages in software development. To shed light on these research questions, we investigate correlations between programming languages and repositories. Furthermore, we look at the popularity of different programming language, and compare the results to the tags on Stack Overflow~\cite{StackOverflow}, the Q\&A website dedicated to programming questions. Even more, we provide a user-driven tree classification of programming languages, and discuss the underlying reasons for this phenomenon. Additionally, we look at the set of programming languages that are used in a repository in more detail. Based on these observation we create the graph of relationship of auxiliary programming languages, and rank them using different metrics such as PageRank, and percentage of lines of code.

Our contributions are summarized as follows:

\begin{itemize}

\item Collection of a large dataset mirroring the state of GitHub, spanning 8 years of interactions (2007-2014); consisting of around 10 million users, 16 million repositories, 11 million user relationships, and over 3 billion contributions. We will make this dataset publicly available to the benefit of the research community.

\item A holistic analysis of GitHub's ecosystem, growth and adoption rate, based on the number of users and repositories.

\item A detailed analysis of the top repositories, and the identification of the popular programming languages on GitHub, and comparing the results with other sources, such as Stack Overflow, and TIOBE~\cite{TIOBE}.

\item Investigation of the correlation between different programming languages, and hierarchical clustering of the different languages and technologies using machine learning techniques.

\item Derive a user-driven phylogenetic tree classification of programming languages based on the user-repository-language interactions. Furthermore, we examine the relationship between the main and auxiliary programming languages in repositories.

\end{itemize}

The rest of the paper is organized as the following. In section~\ref{s:collection} we describe the data collection infrastructure and methodology. Then, in section~\ref{s:analysis} we provide basic analysis and statistics of GitHub's ecosystem, and its growth pattern. In section~\ref{s:pl_rel} we investigate the relationship between programming languages, and their clustering. Followed by an overview of the related work in section~\ref{s:related}. In section~\ref{s:future}, we overview the future work; and finally, we conclude the work in section~\ref{s:conclusion}.

\section{Data Collection: Infrastructure and Methodology}\label{s:collection}

\begin{figure}
\centering
\includegraphics[width=0.5\textwidth]{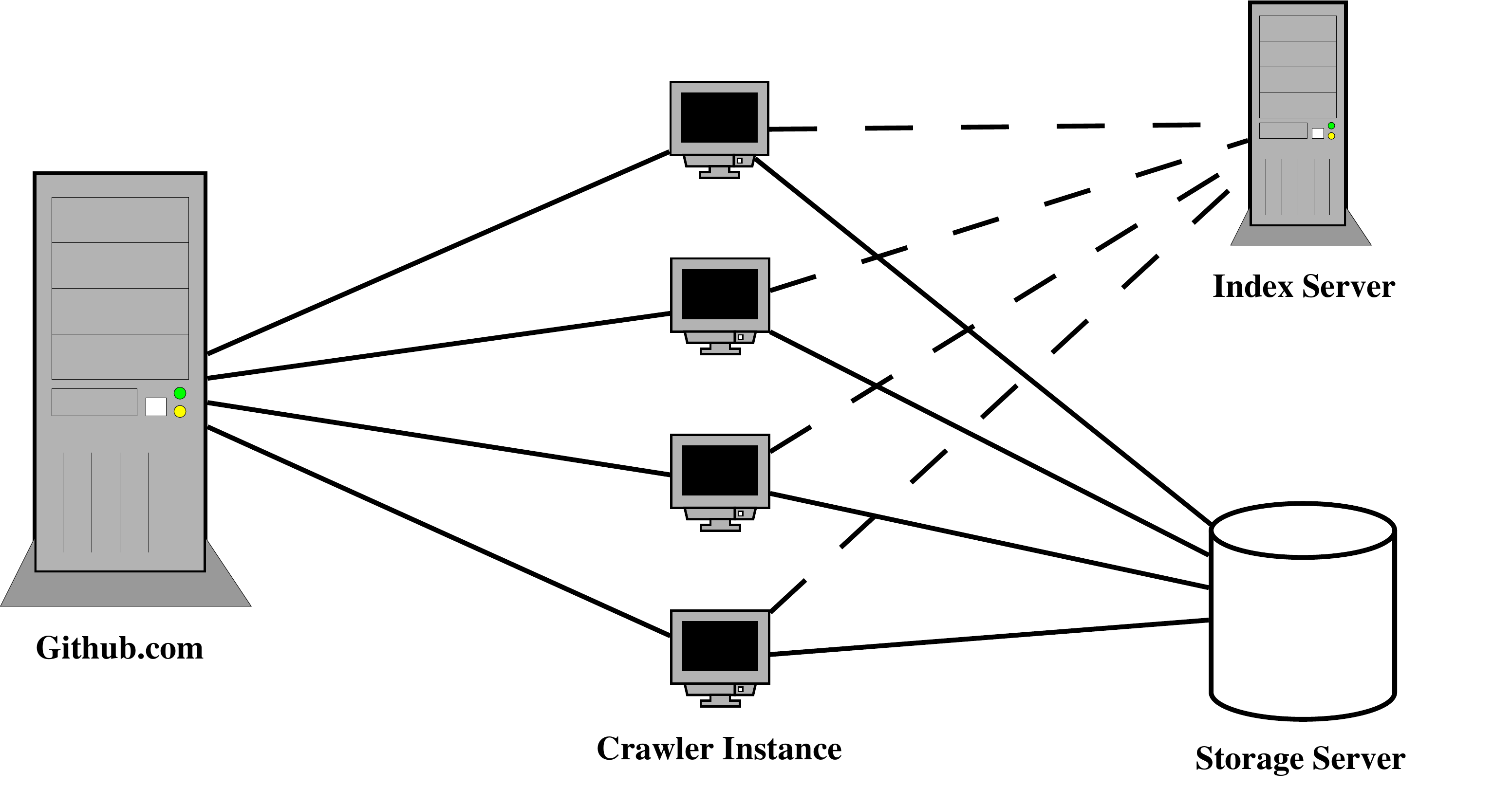}
\caption{Data collection system, with 200 crawler instance, an index server and a storage server. The index server keeps track of the progress of each crawler. The storage server writes the data received from the crawler to persistent storage after parsing it.}
\label{f:data_collection}
\end{figure}

We build our infrastructure to collect and analyze data from GitHub. Although data collections such as GitHub Archive~\cite{gharchive} and GHTorrent~\cite{Gousi13} already exist, for the purpose of our study, these datasets are not sufficient. Both datasets only mirror the public events that happen on GitHub. Therefore, it makes them limited as they do not provide a holistic view of GitHub. We are interested in a holistic study of GitHub. Furthermore, our data set fills the current gap in the aforementioned data collections attempts. For example both GitHub Archive and GHTorrent are available from 2012 onwards. One can extend the snapshot of our collection (as the baseline), to the current state of GitHub, by interpolating the events collected by GitHub Archive and GHTorrent.  The main challenge in collecting the state of GitHub, is the size of data and the rate limit imposed by GitHub. An account and IP address is limited to 5000 queries per hour. Since there are more than 10 million users, 16 million repositories, and 5 million follower-followee relationships, collecting this data with the rate limit would take months. We implemented and deployed a resilient distributed collection system, consisting of 200 collection advantage points. This is a one time data collection, and to avoid introducing substantial load on GitHub's server, we geographically spread the data collection advantage points. We also implemented a backoff mechanism to address the load issues. The following describes the data collection infrastructure, methodology and the data that we collected.

\begin{figure}
\centering
\includegraphics[width=0.5\textwidth]{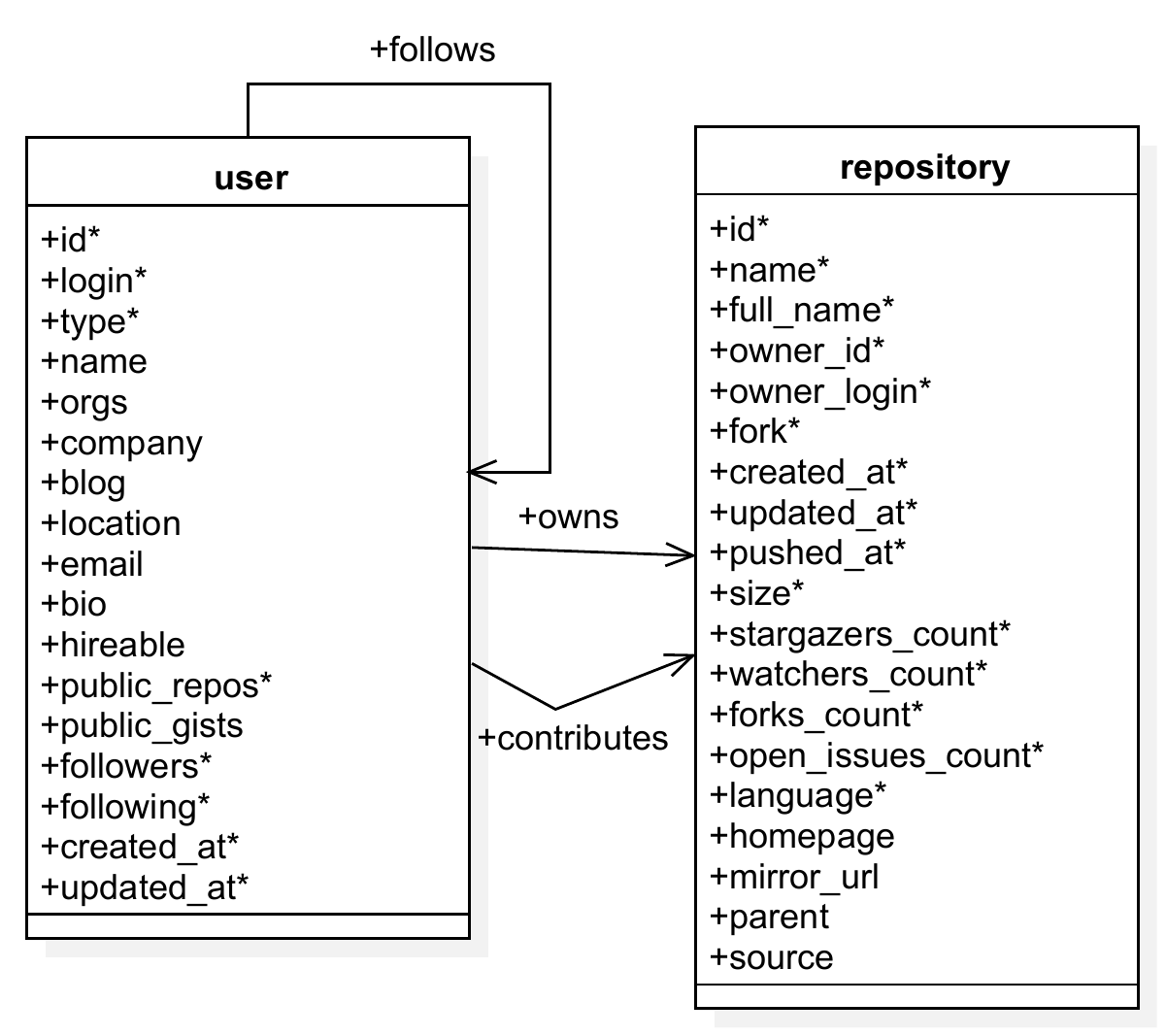}
\caption{The attributes collected for each entity (user and repository), and the relationship between users and repositories. A user can own and contribute to repositories. Additionally, a user can follow another user. Note that the attributes with a star must have a value.}
\label{f:data_rel}
\end{figure}

GitHub provides a RESTful API (currently version 3) to make queries about the users and repositories. The list of current users can be queried at \emph{https://api.github.com/users}, and the list of current repositories is accessible at \\ \emph{https://api.github.com/repositories}. More detailed information about users and repositories can be obtained by making specific GET queries for a repository or a user. The result is returned in JSON format. Queries make use of paging to limit the amount of data that is returned from the server. Figure~\ref{f:data_collection}, shows a high-level description of our infrastructure. We divided the number of users and repositories between our 200 data collection advantage points. Each client is responsible for collecting a range of IDs. We use free public cloud system, but unfortunately the free tier systems do not offer persistent storage and the clients can crash or reboot. To address this limitation we adapt our system to make it resilient to such failures, by implementing an index server and a storage server. The index server keeps track of the current ID for each data collection client. We implemented an in memory key-value storage system that provides a RESTful API. At the beginning, when a client starts, it queries the index server for the last ID it has collected the data for. We use the IDs to keep track of the users and repositories since the login names and the repository names can change, however IDs stay the same. Furthermore, IDs increment in order of user/repository creation time. The IDs of the deleted users and repositories do not get reused, therefore there can be gaps in the IDs. The clients parse the JSON data received from GitHub and collect the attributes for the users and the repositories. As mentioned earlier, we implemented a number of ad-hoc techniques to avoid overloading the GitHub's servers. For example, we introduce delayed request sequences, and backoff mechanisms if the average response time from the server exceeds the acceptable threshold. In the next step the clients send the parsed data to the persistent storage system. Since the server receives a large amount of traffic from 200 clients, it keeps a queue for the incoming data and a worker processor reads the data in the queue, and writes it into the persistent storage. We serialize the data and write it to a flat file. Later, in the processing stage, another script would create a dataset based on the arguments that we need to analyze. Using this mechanism allows us to avoid loading the entire data set into the volatile memory. During the processing stage records are loaded and evicted linearly. We only collect the main attributes, and we infer the other attributes based on the collected data. For users, we collect \texttt{name}, \texttt{ID}, number of \texttt{public repositories} and \texttt{creation time} among other features. For repositories, we store metadata such as the \texttt{owner}, \texttt{owner}, number of \texttt{stars} and \texttt{creation time}. Figure~\ref{f:data_rel} depicts the features that we collected for users and repositories and the relationship between the entities. Note that the attribute with a star, must have a value. The remaining attributes are free text entries and can be empty.
\section{Analysis of GitHub's Ecosystem}\label{s:analysis}

\begin{figure*}
\centering
\includegraphics[trim={0 0.6cm 0 0.6cm}, height=9cm,width=18cm]{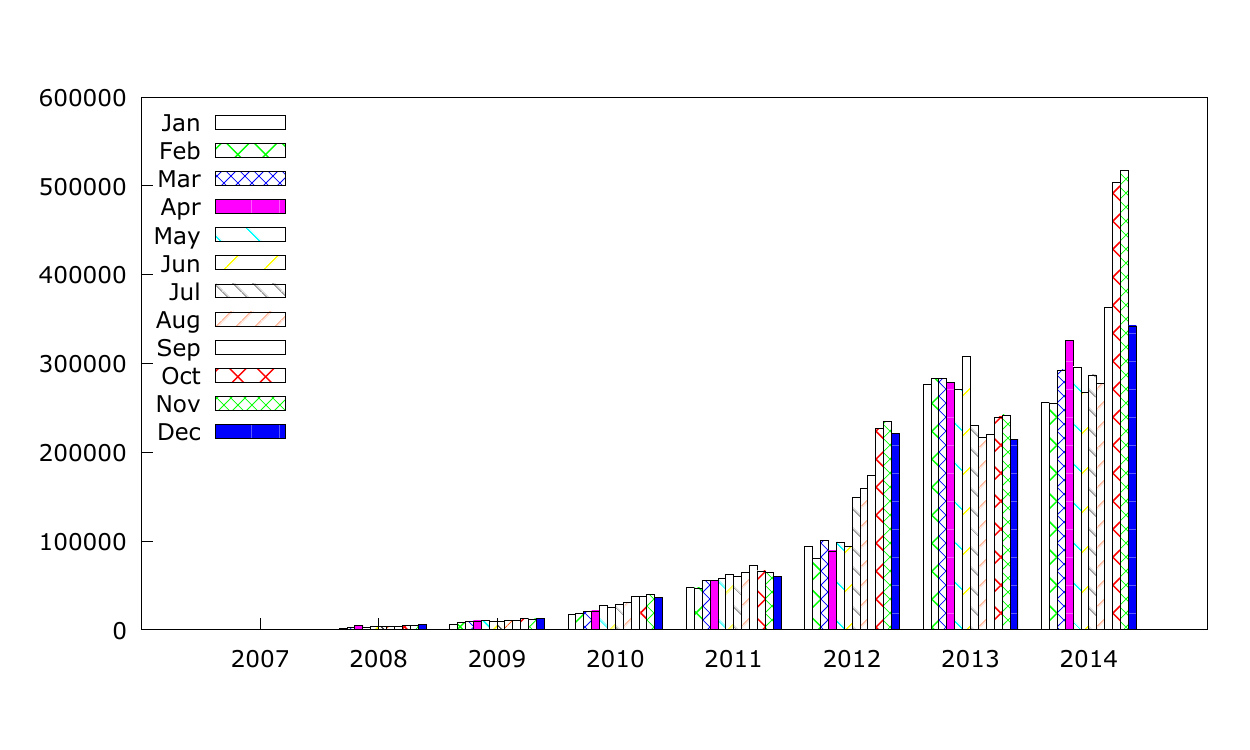}
\caption{Histogram of creation of users on GitHub. As we can see, the number of users has been increasing exponentially until 2014. For instance, the number of users created in 2013 and 2014, is twice more than the number of users created from 2007 to 2012, combined. However, 2014 is an inflection point in GitHub’s growth.}
\label{f:hist-user-creation}
\end{figure*}

In this section we start by a discussion of macro-statistics of GitHub. Later, in the following section we delve deeper into different aspects of the aforementioned ecosystem. In our dataset there are around 10 million (9993767) users and around 17 million (16812452) repositories. Figure~\ref{f:hist-user-creation}, shows the histogram of users joining GitHub, at different months from 2007 to 2014. As we can see, the number of users has been increasing exponentially until 2014. For instance, the number of users created in 2013 and 2014, is twice more than the number of users created from 2007 to 2012, combined. However, it seems like 2014 is an inflection point in GitHub's growth, and the number of newly registered users in 2014, is only 30\% more than 2013. This trend follows the diffusion of innovation~\cite{rogers2010diffusion} principle, and the late majority are joining now. The large majority (about 95\%, 9533220 out of 9993767) of these accounts are ``Users'', and the rest (460547), are ``Organizations''.

As expected the number of users registering on GitHub in December is lower than the previous two months, because of the holidays. The sudden increase in the number of new users in 2012, can partially be explained by non technical events and the press coverage that GitHub received in 2012. For example the co-founders PJ Hyett and Chris Wanstrath were named 30 under 30 by Forbes~\cite{forbes30}, GitHub won the ``Best Overall Startup'' award by TechCrunch~\cite{techcrunch} and was selected in the Forbes' top 10 startups~\cite{forbes10}. Furthermore, in 2012 Andreessen Horowitz (4 billion dollar venture capital firm founded by Marc Andreessen and Ben Horowitz), invested 100 million dollars in GitHub, which was also the first ever outside investment~\cite{andersonhorrowitz}. As we can also see from the data, 2012 has been a critical year in GitHub's success and popularity. Our data set shows how such socioeconomic external factors had impact on the growth and success of GitHub.

\begin{figure}
\centering
\includegraphics[width=0.5\textwidth]{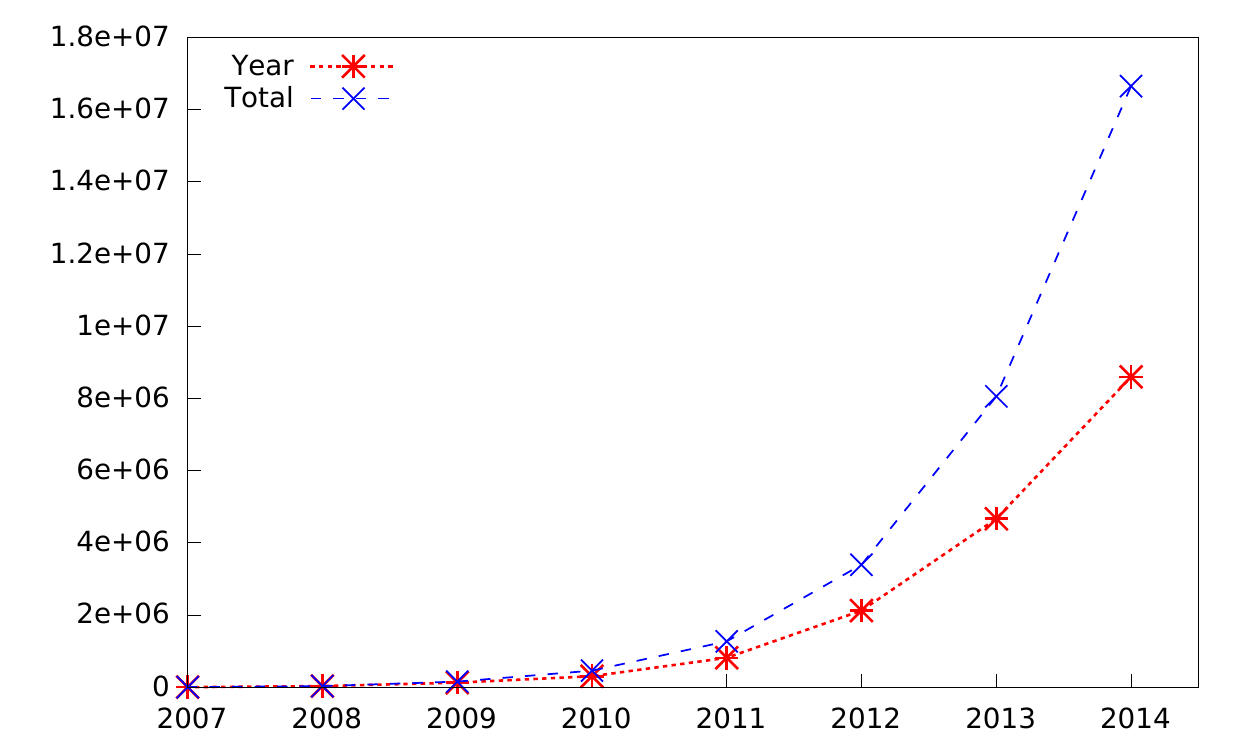}
\caption{Growth of the repositories in GitHub. The dashed blue line shows the cumulative number of repositories, while the red line indicates the number of repositories created in each year.}
\label{f:repo-growth}
\end{figure}

The repositories are written in more than 220 different programming languages. Note that these programming languages are based on GitHub's definition of programming languages. For instance, \texttt{Makefile} and \texttt{Batchfile} are also considered languages in GitHub's definition. The top 5 programming languages in terms of number of repositories, in order are: Java Script, Java, Ruby, Python, and PHP. As Figure~\ref{f:repo-growth} depicts, the exponential growth rate of the repositories. The number of repositories created in years 2007 to 2012, combined, is less than number of repositories created in 2013, and less than half of the repositories created in 2014. This observation is in harmony with the observation on the adoption of GitHub by the users. More than 55\% of the repositories are original (7304258 repositories are fork and 9508194 are original). Meaning these are not a direct fork of another repository on Gihub, though they can be a re-upload of another repository. Only 8603 of the forked repositories outshine their source repository, in terms of number of stars. Meaning a fork having more stars than its source. However, the gap is marginal and not significant. The distribution of the stars is uneven and follows the power law characteristics, where a very small fraction takes the majority of resources. For example, 80\% of the repositories have no stars, and 99\% of the repositories have 13 stars or less. Even, after excluding the projects with no stars, still, 95\% of the remaining repositories have 13 stars or less.

To shed light on the nature of GitHub's programming languages, we analyze repositories in more detail in the following section.

\begin{table*}[t]
  \centering
  \begin{tabular}{| c | c | c | c | c | c | c | c | c |}
    \hline
    Programming Language& Popularity   		& Stars   & Commit 		& Size  	& N Repos  			   & Fork 		& TIOBE 	& StackOverflow \\ \hline
    Java Script 		& 1 				& 7328824 & 3 			& 3 		& 1			 		   & 1			& 6			& 10			\\ \hline
    Ruby				& 2 				& 2997381 & 7 			& 8 		& 3			 		   & 3			& 18		& 14			\\ \hline
	Python 				& 3 				& 2387965 & 6 			& 5 		& 4			 		   & 4			& 8			& 20			\\ \hline
	Objective-C 		& 4 				& 1905905 & 11 			& 10 		& 9			 		   & 7			& 4			& 22			\\ \hline
	Java 				& 5 				& 1854823 & 4 			& 1 		& 2			 		   & 2			& 1			& 1				\\ \hline
	PHP 				& 6 				& 1566619 & 5 			& 6 		& 5			 		   & 5			& 7			& 28			\\ \hline
	CSS 				& 7 				& 1172607 & 8 			& 9 		& 6			 		   & 9			& -			& 30			\\ \hline
	C					& 8 				& 1127017 & 1 			& 2 		& 7			 		   & 6			& 2			& 23			\\ \hline
	C++					& 9 				& 877088  & 2 			& 4 		& 8			 		   & 8			& 3			& 3				\\ \hline
	Go 					& 10 				& 580677  & 12 			& 23 		& 14			 	   & 13			& -			& $>$ 200		\\ \hline
  \end{tabular}		
  \caption{Programming Languages popularity using different metrics. Java Script is at the first place (overall number of stars) by a far margin.}
  \label{t:pl-rank}
\end{table*}

\subsection{Delving into the Repositories} \label{ss:repos}

In this section we investigate what are the top 10 programming languages, by interpolating the number of stars for repositories in GitHub. Stars in GitHub are equivalent of ``like'' in social platforms such as Facebook. The other metrics that we look at our the overall number of commits, size of software artifacts repositories (bytes), overall number of repositories, and overall number of forks. We also compare the results to the rankings provided by TIOBE and Stack Overflow. As we can see in Table~\ref{t:pl-rank}, Java Script is by a far margin on the top of the list followed by Ruby and Python, even though Java is the top programming language according to Stack Overflow and TIOBE. Note that Stack Overflow is based on the number of questions that are asked in a different programming language, meanwhile, we are using a popularity measure. It is not surprising that Java is ranked number one according to Stack Overflow, given that simply reading from a file requires substantially more lines of code, compared to e.g., Python (1 line of code).

\begin{figure*}
\centering
\includegraphics[height=8.7cm,width=18cm]{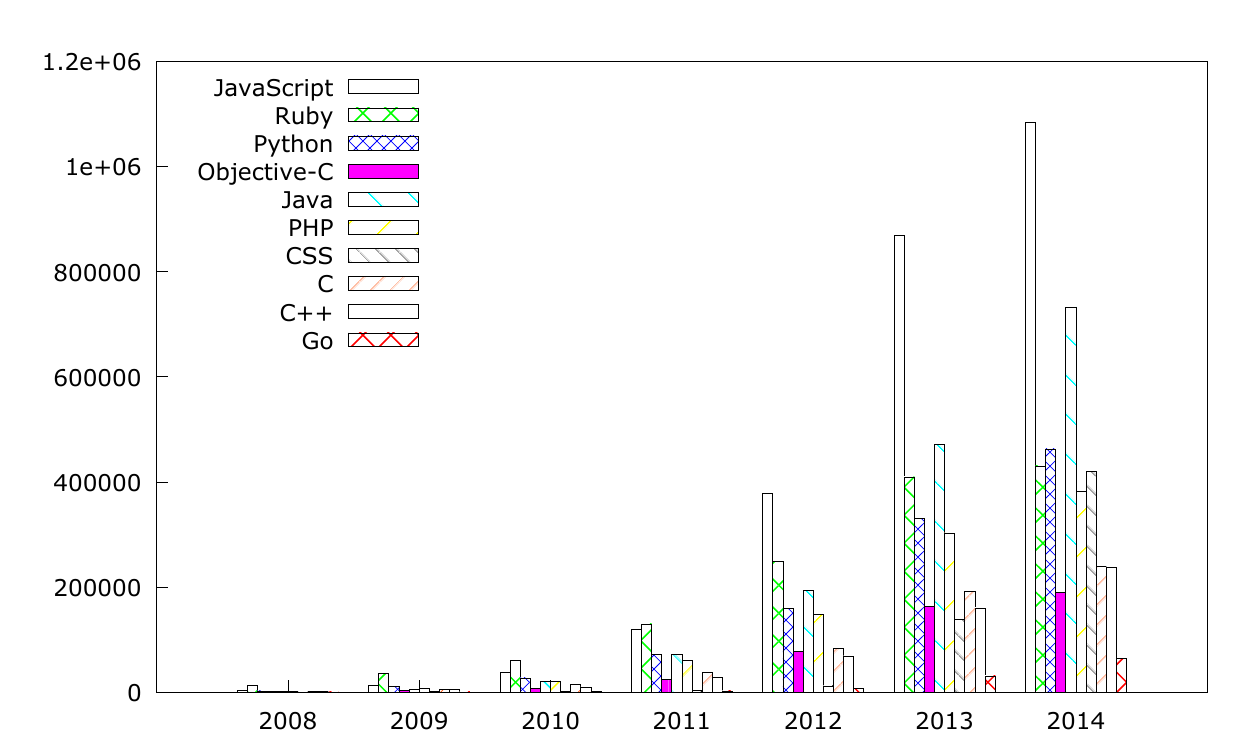}
\caption{Histogram of repository creation year for the top 10 popular programming languages. Until 2011 Ruby was at the first place in terms of the number of repositories, and after Java Script takes the first place.}
\label{f:hist-create-repos}
\end{figure*}

The other top 3 programming language that are not in our top 10 popular languages are: C\#, Perl and R. The aforementioned languages have the following ranks in our table. C\# is at 12th place, followed by Perl at 17th place, and finally R at 25th place. C\# is mostly restricted to .NET framework and Microsoft; Perl is gradually being replaced by other scripting languages; and R is a statistical domain specific language, and recently is being challenged by general programming languages such as Python.

Another question is whether the top 10 programming languages appeared on GitHub at the same pace? As we can see in Figure~\ref{f:hist-create-repos}, until 2011 Ruby was the dominant programming language on GitHub, in terms of the number of projects. Even more, in 2007, the very first and only repository on GitHub (Grit) was in Ruby. Grit, is written by GitHub's co-founder, Tom Preston-Werner. Java Script, Ruby and Python are the top 3 programming languages until 2012. Java takes the second place in 2013 and 2014, after Java Script. The rise of Java can be attributed to the popularity of Android platform and wearable devices, and third party libraries. The abundance and utility of third parties is a deciding factor in the success and adoption of programming languages. According to Gartner~\cite{gartner}, the share of Android devices raised from 30\% in 2010 to 84\% in 2015.

By looking at the contributions to repositories we find out that, the repositories with the most number of commits are generated by bots to change the appearance of the user's timeline, or participate in a contest. This is an interesting phenomenon worth investigating further, to see whether these are malicious or benign activities. However, the first real repository is \texttt{google/capsicum-linux}, which also has the most number of contributors, with over 4500 contributors. As a matter of fact, all top 5 repositories in terms of the number of contributors are Linux kernel related, and two of which are owned by Google.

\subsection{An Inquiry of Top Repositories}\label{ss:top_repo}

To have a better understanding of the repositories on GitHub, and specially what each programming language is mostly used for, we look at the top 5 repositories. We select the repositories based on the number of stars for our top 10 programming languages. Here we discuss a summary of these findings. The most prominent project written in the Go programming language is Docker. Docker is a suite to automate the deployment of applications inside software containers, and was released on GitHub in 2013. It is nearly twice as popular as the second repository. On the contrary, to the other programming languages, such as Java Script, Ruby or Python, the top repository have a close popularity measure. The top 5 repositories in Java Script were released between 2009 and 2010. The first repository has around 34000 stars, and the 5th one has around 28000 stars. Furthermore, the projects show a closer relationship between their category and classification. For instance 4 out of the top 5 repositories in Ruby, are in the web application/framework category. As expected, the more domain specific programming languages such as PHP, CSS, and Java Script, are largely used in web based projects. Another example is Objective-C, which is entirely used in iOS/Mac OS X based project. However, Java user base is mostly spread between two communities, Android development, and distributed big data based project. An interesting and challenging future work is the analysis of the growth and adoption of languages such as Java, in different communities/software suites.

By looking at the projects' release timeline on GitHub, we find out that Go projects are more recent compared to other programming languages. This can be an indication of the fact that now Go has reached a maturity level (e.g., in terms of libraries ecosystem) that makes it a suitable choice for open source projects. On the other hand, Ruby, Java Script and C projects are older. As mentioned earlier, Ruby was the first programming language to appear on GitHub, and until 2011 had been the dominating programming language, in terms of number of projects (Figure~\ref{f:hist-create-repos}). The oldest popular projects (created in 2008) are: Jekyll, and Rails. Jekyll, is a simple, blog-aware, static site generator for personal, project, or organization sites, written in Ruby; Co-authored by GitHub's co-founder. Rails, is a web-application framework to create database-backed web applications in Ruby.

\section{Programming Languages \\ Relationship}\label{s:pl_rel}

\begin{table*}[t]
  \centering
  \begin{tabular}{| c | c | c | c | c | c | c | c | c | c | c |}
    \hline
     & Java Script & Python  & Ruby   & Objective-C  & Java  & PHP  & CSS 	& C & C++ & Go 			\\ \hline
    Java Script 		& - & 4.29 & 4.40 & 1.00 & 7.93 & 5.66 & 20.77 & 6.81 & 3.77 & 0.64			\\ \hline
    Python				& 10.27 & - & 1.82 & 0.64 & 9.27 & 2.07 & 15.76 & 17.78 & 9.28 & 1.19		\\ \hline
	Ruby 				& 13.76 & 2.37 & - & 1.72 & 7.10 & 2.30 & 25.76 & 8.76 & 4.22 & 1.33		\\ \hline
	Objective-C 		& 7.30 & 1.97 & 4.02 & - & 8.80 & 1.50 & 11.29 & 9.00 & 5.15 & 0.56			\\ \hline
	Java 				& 6.58 & 3.22 & 1.89 & 1.00 & - & 2.39 & 5.91 & 4.90 & 3.23 & 0.29			\\ \hline
	PHP 				& 12.21 & 1.87 & 1.59 & 0.44 & 6.23 & - & 17.16 & 5.43 & 2.99 & 0.39		\\ \hline
	CSS 				& 19.78 & 6.28 & 7.85 & 1.47 & 6.78 & 7.57 & - & 3.65 & 2.19 & 0.87			\\ \hline
	C					& 11.78 & 12.86 & 4.85 & 2.13 & 10.21 & 4.34 & 6.63 & - & 12.48 & 1.66		\\ \hline
	C++					& 8.11 & 8.35 & 2.91 & 1.51 & 8.36 & 2.98 & 4.95 & 15.53 & - & 0.77			\\ \hline
	Go 					& 13.75 & 10.67 & 9.08 & 1.63 & 7.56 & 3.90 & 19.56 & 20.55 & 7.66 & -		\\ \hline
  \end{tabular}		
  \caption{The correlation between our top 10 Programming Languages. The correlations are calculated over the data collected from the commits that users made to different repositories.}
  \label{t:pl-corr-commit}
\end{table*}

\begin{table*}[t]
  \centering
  \begin{tabular}{| c | c | c | c | c | c | c | c | c | c | c |}
    \hline
     & Java Script & Python  & Ruby   & Objective-C  & Java  & PHP  & CSS 	& C & C++ & Go 			\\ \hline
    Java Script 		&- & 4.46 & 5.34 & 1.42 & 8.00 & 6.83 & 14.63 & 9.59 & 4.10 & 0.55			\\ \hline
    Python				&14.86 & - & 2.34 & 0.97 & 11.75 & 2.61 & 16.06 & 30.24 & 11.85 & 1.30		\\ \hline
	Ruby 				&18.54 & 2.44 & - & 1.75 & 9.62 & 2.76 & 20.31 & 18.14 & 5.94 & 1.12		\\ \hline
	Objective-C 		&10.39 & 2.13 & 3.68 & - & 10.36 & 1.80 & 9.86 & 20.39 & 6.91 & 0.54		\\ \hline
	Java 				&8.54 & 3.76 & 2.95 & 1.51 & - & 2.93 & 5.38 & 8.62 & 3.75 & 0.34			\\ \hline
	PHP 				&18.23 & 2.09 & 2.12 & 0.66 & 7.34 & - & 16.37 & 10.84 & 4.36 & 0.51		\\ \hline
	CSS 				&22.32 & 7.35 & 8.92 & 2.06 & 7.70 & 9.36 & - & 6.49 & 2.70 & 0.84			\\ \hline
	C					&15.80 & 14.96 & 8.61 & 4.60 & 13.32 & 6.69 & 7.01 & - & 14.10 & 2.39		\\ \hline
	C++					&10.87 & 9.43 & 4.53 & 2.51 & 9.33 & 4.33 & 4.70 & 22.69 & - & 0.90			\\ \hline
	Go 					& 17.74 & 12.66 & 10.44 & 2.41 & 10.31 & 6.15 & 17.80 & 47.03 & 11.00 & -	\\ \hline
  \end{tabular}		
  \caption{The correlation between our top 10 Programming Languages. The correlations are based on the repositories created by different users.}
  \label{t:pl-corr-repo}
\end{table*}

In this section, we look at the relationship between the top 10 programming languages on GitHub. Our goal is to find out whether there is a correlation between programming languages. For example, is someone who codes in Python more likely to also code in Ruby, or PHP? The answer to this question can help us to understand the fundamental question of what is the general relationship between different programming languages. Are they only an evolution of the older generation of programming languages, or whether different technologies try to answer and solve a different set of problems. Of course, all of programming languages that we are considering are Turing complete, with the exception of CSS, but do people use them in the same way and for the same purposes, or each one is more suitable for a different set of problems. This suitability can be because of the innate features of the programming language, or an external factor. Namely, a killer application/library for that language, or adoption of the language in a certain product. For example, iOS and Mac OS X's Software Development Kits (SDK) are based on Objective-C. As a results all top 5 repositories in Objective-C are Apple, and more specifically iOS/Mac OS related.

To calculate the correlations between programming languages, we look at the commits that are made by users (Table~\ref{t:pl-corr-commit}) and the repositories that are created by each user (Table~\ref{t:pl-corr-repo}). We use the following methodology to calculate the values. Imagine, Alice codes in Python, Ruby, and C; Bob codes in Python, and C; and Carol codes in Python, and Ruby. Therefore, the correlation between Python and Ruby, is ${2 \over 3} = 0 .66 = 66\%$ (${Alice + Carol \over Alice + Bob + Carol}$), and the correlation between Ruby and Python is 1 = 100\% (${Alice + Carol \over Alice + Carol}$). Meaning, 66\% of people who use Python also use Ruby, while 100\% of users who use Ruby also use Python. Note that these relationships are asymmetric. As we can see, this translate to the conditional probability, i.e., $P(A|B)$. The probability that A occurs given B.

\begin{align*}
P(A|B) = {P(A \cap B) \over P(B)}
\end{align*}

\begin{align*}
P(Python|Ruby) = {P(Python \cap Ruby) \over P(Ruby)}
\end{align*}

As we can see in Table~\ref{t:pl-corr-commit}, Ruby programmers are most likely to also code in Java Script or CSS rather than in Go or Objective-C. Such relationships can be linked to the popularity of web frameworks such as Ruby on Rails, and Sinatra. Note that projects categorized as CSS, can in fact be Java Script frameworks where the majority of the code base is CSS files. Such usage patterns also surfaces in other programming languages as well. PHP programmers are much more likely to use Java Script and CSS than any other programming language. Considering that the aforementioned programming languages are mostly used in web development, such a closed relationship is expected. On the other hand, Python and Go programmers are much more likely to program in C than other programming languages. The de facto implementation of these programming languages is in C, and are more likely to be used in systems related software and projects, compared to Ruby or Java Script. 

As we can see in Table~\ref{t:pl-corr-repo}, the same observations hold for the data based on the programming language of the repositories that are created by users, as opposed to the repositories that users have contributed to. Such correlation and entangled behavior, motivate us to further investigate the clustering behavior of programming languages, by using unsupervised machine learning algorithms.

\subsection{Clustering of Programming Languages}\label{ss:pl_cluster}

\begin{figure*}
        \centering
        \begin{subfigure}[b]{0.48\textwidth}
                \includegraphics[width=\textwidth]{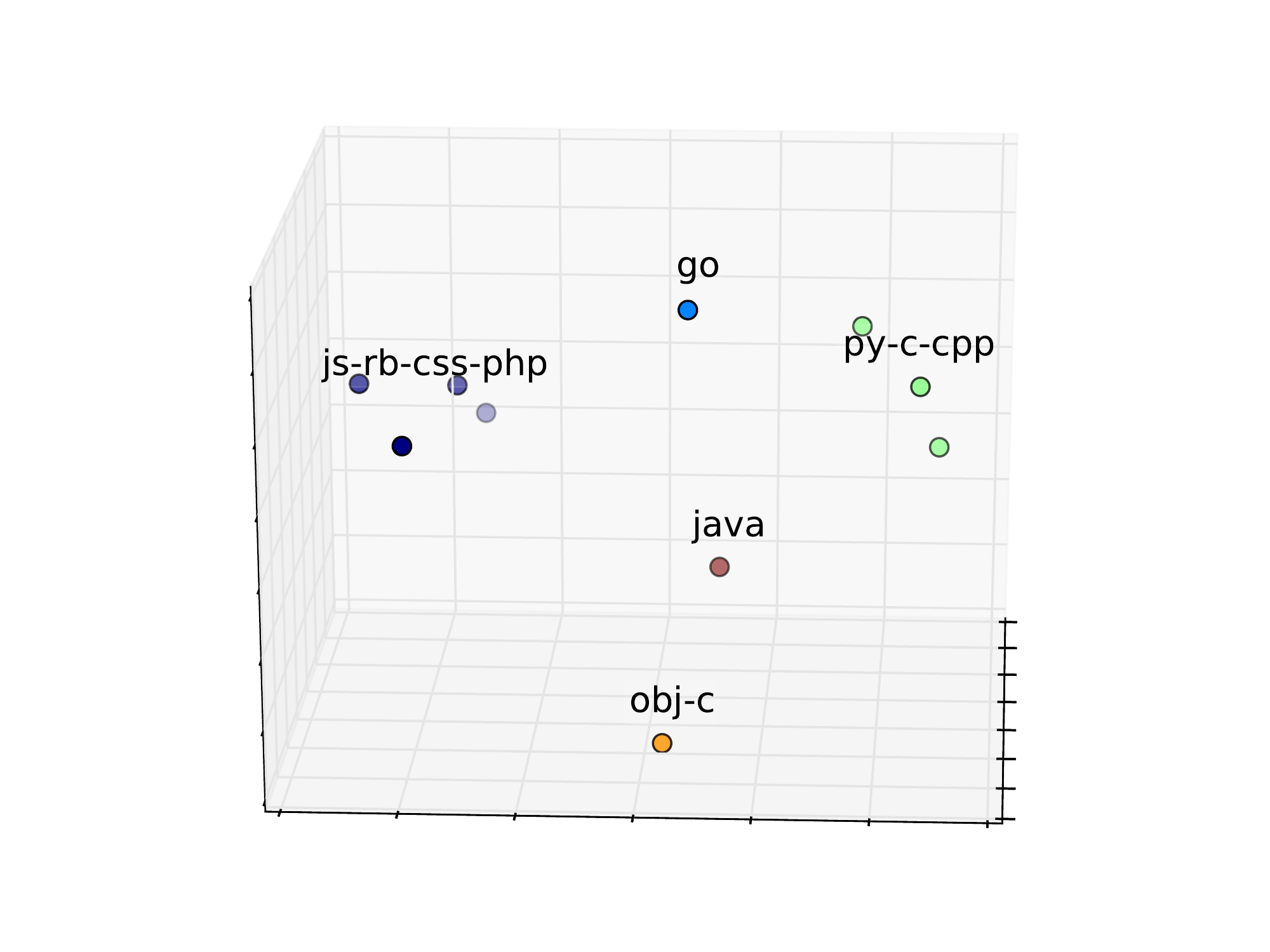}
                \caption{5 clusters of programming languages, based on the commits (Table~\ref{t:pl-corr-commit}); the web programming languages (Java Script, Ruby, PHP, CSS), system oriented languages 					(Python, C, C++), Android (Java), iOS/Mac OS X (Objective-C) and “trending/upcoming” (Go).}
                \label{fig:5_commit}
        \end{subfigure}%
        ~~~~~~~\quad 
        \begin{subfigure}[b]{0.48\textwidth}
                \includegraphics[width=\textwidth]{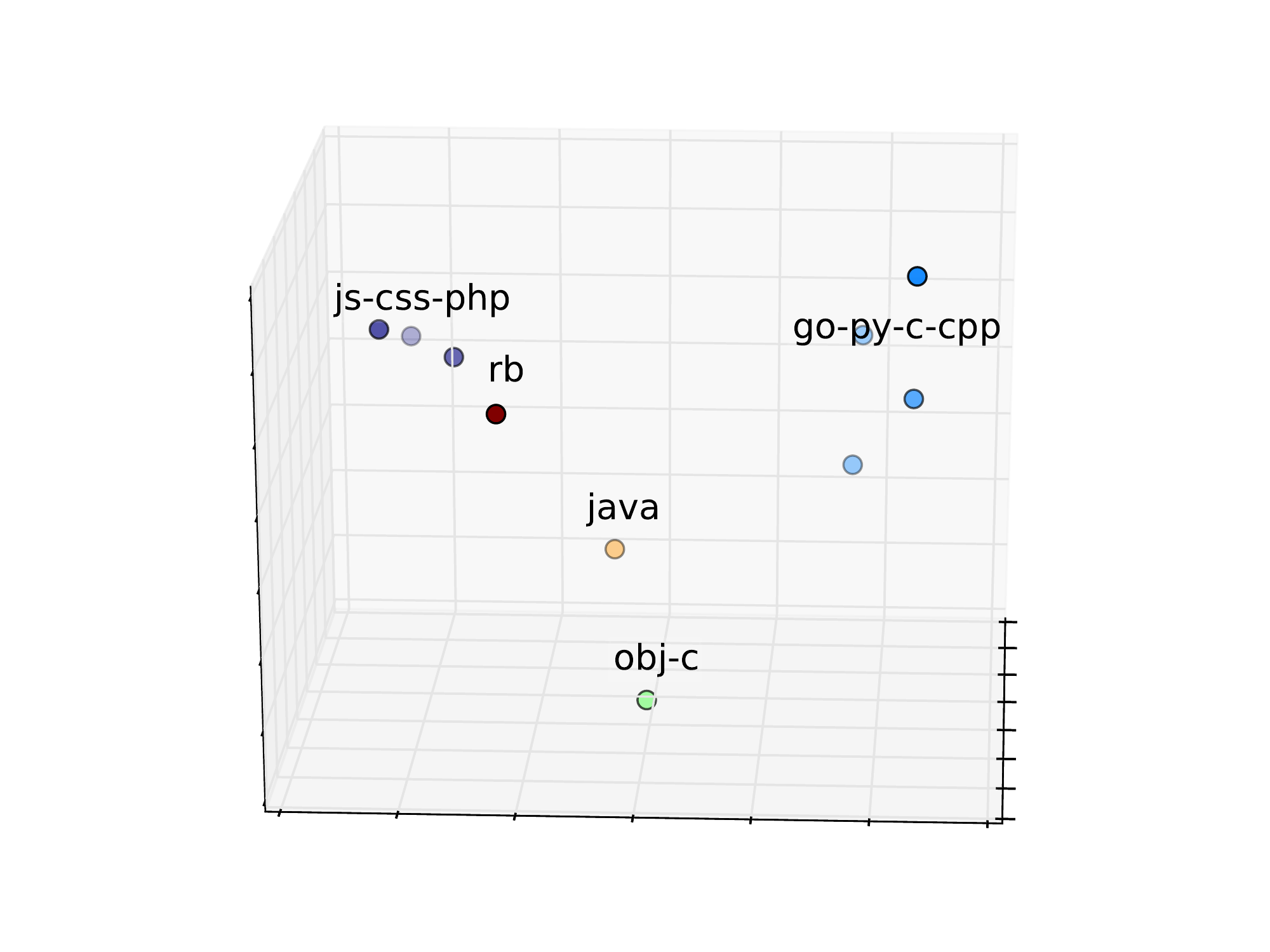}
                \caption{5 clusters of programming languages, based on users' repositories (Table~\ref{t:pl-corr-repo}); the web programming languages (Java Script, PHP, CSS), system oriented languages (Python, Go, C, C++), Android (Java), iOS/Mac OS X (Objective-C) and Web/Minimalist (Ruby).}
                \label{fig:5_repos}
        \end{subfigure}

        \caption{Clustering of programming languages}\label{fig:pl_cluster}
\end{figure*}

In the following we cluster the top 10 programming languages on GitHub, using k-means~\cite{Arthur:2007:KAC:1283383.1283494} clustering. K-means is a vector quantization method originated from signal processing, which is used for clustering. We perform a maximum of 300 iterations with 10 centroid seeds. We use the values in Table~\ref{t:pl-corr-commit} and Table~\ref{t:pl-corr-repo} as our input dataset. Each column represents a feature and each row is an entry. Figure~\ref{fig:5_commit} and Figure~\ref{fig:5_repos} depict the clustering of programming languages. To be able to visualize the data, we perform a dimension reduction (from 10 dimensions to 3 dimension), using principal component analysis (PCA)~\cite{jolliffe2002principal}. PCA is an orthogonal transformation to convert a set of correlated variables into a set of values of linearly uncorrelated variables. As we can see in Figure~\ref{fig:5_commit}, there are 5 clusters of programming languages; the web programming languages (Java Script, Ruby, PHP, CSS), system oriented (Python, C, C++), Android (Java), iOS/Mac OS X (Objective-C) and ``trending/upcoming'' (Go). Furthermore, if we look at the repositories created by users we see slightly different results. As we can see in Figure~\ref{fig:5_repos}, the web programming languages still consist of Java Script, PHP, and CSS. However, with 5 clusters, Ruby emerges as its own cluster, and Go falls within the system oriented programming languages, alongside Python, C, and C++. Clustering of Java Script, PHP, and CSS is expected, as previous measurements~\cite{Netcraft} have estimated 39\% of the backend webservers were running PHP in 2013. However, emergence of Ruby in the web programming languages is more specific to the GitHub platform. As mentioned before Ruby was the first and the dominant programming language on GitHub until 2011. Among the  big web companies that used Ruby for their back-end development is Twitter~\cite{twitterlang}. Based on these observations we look at the gradual separation of programming languages, and create a user-driven tree classification of programming languages.

\begin{figure}
\centering
\includegraphics[width=0.5\textwidth]{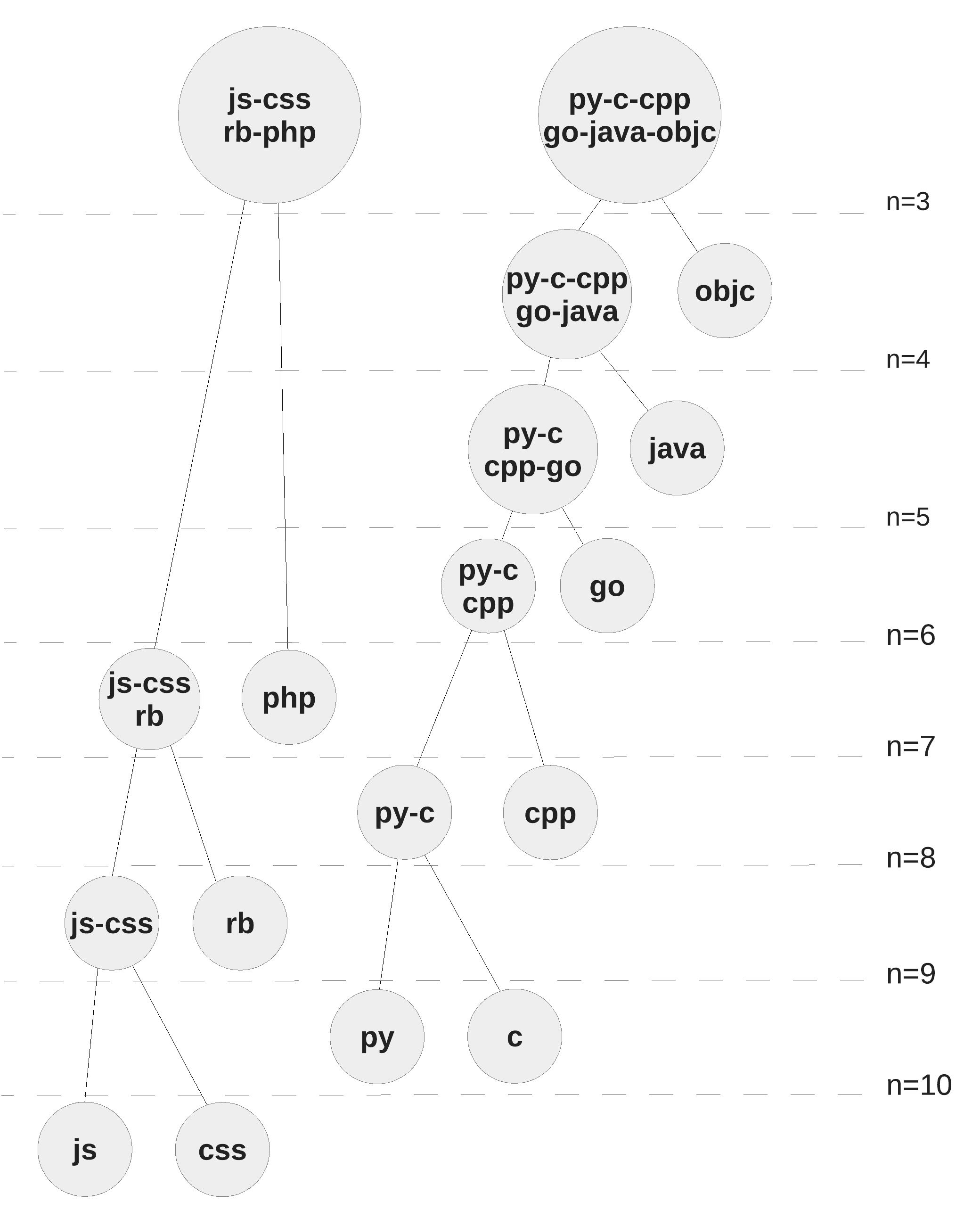}
\caption{Hierarchical phylogenetic clustering and separation of programming languages. At first, there are two families of programming languages; the web programming languages and the ``others''.}
\label{f:lang-cluster-gradual}
\end{figure}

We start with two clusters and gradually increase the number of clusters. Figure~\ref{f:lang-cluster-gradual}, is based on the commits data (Table~\ref{t:pl-corr-commit}). As we can see, at first there are two families of programming languages; the web programming languages (Java Script, Ruby, PHP, CSS), and the ``others'' (Python, C, C++, Go, Java, Objective-C). The first programming languages that separates from the herd of ``others'', is Objective-C, which has a different use and platform from the rest (almost exclusively iOS/Mac OS X). Next, Java emerges as a different category (Android, distributed Big Data), followed by Go. However, in the next step PHP leaves the web family, followed by Ruby. The last separation of ``other'' family is between Python and C (Python's reference implementation, CPython, is in C). Java Script and CSS, keep their close tie, until the last clustering step. Note that, as mentioned earlier many of the CSS repositories are in fact Java Script libraries where the majority of code base is CSS.

\subsection{Investigation of Auxiliary Programming \\ Languages}\label{ss:top100}

As mentioned earlier a project may be written in more than one programming language. In the previous section we considered the dominant programming languages of a repository, in terms of lines of code. In this section we look further into all the programming languages that are used in a repository. To investigate a good representative of the open source project we look at the top 100 repositories for each of the top 10 programming languages, a total of 1000 repositories.

\begin{table}[t]
  \centering
  \begin{tabular}{| c | c |}
    \hline
    PageRank Ranking	& Programming Language  \\ \hline

   	1	&	HTML			\\ \hline
    2	&	Perl			\\ \hline
	3	&	Makefile		\\ \hline
	4	&	Batchfile		\\ \hline
	5	&	Shell 			\\ \hline
	6	&	CoffeeScript 	\\ \hline
	7	&	JavaScript 		\\ \hline
	8	&	Python			\\ \hline
	9	&	Ruby			\\ \hline
	10	&	PHP				\\ \hline
  \end{tabular}		
  \caption{Top 10 Programming Language, based on the PageRank algorithm. As we can see, HTML is used in many projects and is used with other programming languages, followed by scripting languages to facilitate tasks in the projects.}
  \label{t:top100-10pgrank}
\end{table}

We represent the relationship between the programming languages as a directed graph. The nodes of the graph represent the programming languages, and we consider an edge between two nodes if the corresponding programming languages are used in a project. The source of the edge is the dominant programming language and the destination is the other programming programming language. For example, if a repository is written in Python, Ruby, C and Java, where the majority of the source code is in Python, we consider edges from Python to Java, Python to Ruby, and Python to C. To define it more formally, $|V|$ = set of all programming  languages, $|E|$ = set of all edges. Therefore, $\forall u, v \in V$, $\exists (u,v) \in E$, $\iff$ $u$ and $v$, are used in a same project, where $u$ is the main, and $v$ is the auxiliary language.

After forming such a graph we use the PageRank~\cite{page1999PageRank} algorithm to rank the programming languages. It is an algorithm used by Google to determine the importance of the websites, based on the other websites that link to it. PageRank works in multiple iterations, and in the first iteration it assigns the same score to all nodes. In the future iterations it updates the scores, where the score of node $p$ is divided between the other nodes it points to. Therefore each node receives $1\over L(p)$ of the score. $L(p)$ is the number of outbound links from $p$. Therefore, the following formula calculates the score of page $p$ at iteration $i$:

\begin{align*} \label{pgrankeq}
PR(p_i) = {1 - d \over N} + d \sum\limits_{p_j \in M(p_i)}{PR(p_j) \over L(p_j)}
\end{align*}
where $p_i$ is the node under consideration, $M(p_i)$ is the set of nodes that link to $p_i$, $L(p_j)$ is the number of outbound edges from node $p_j$, and N is the total number of nodes.

\begin{figure*}
\centering
\includegraphics[trim=4cm 0 -65 0cm, width=1.2\textwidth]{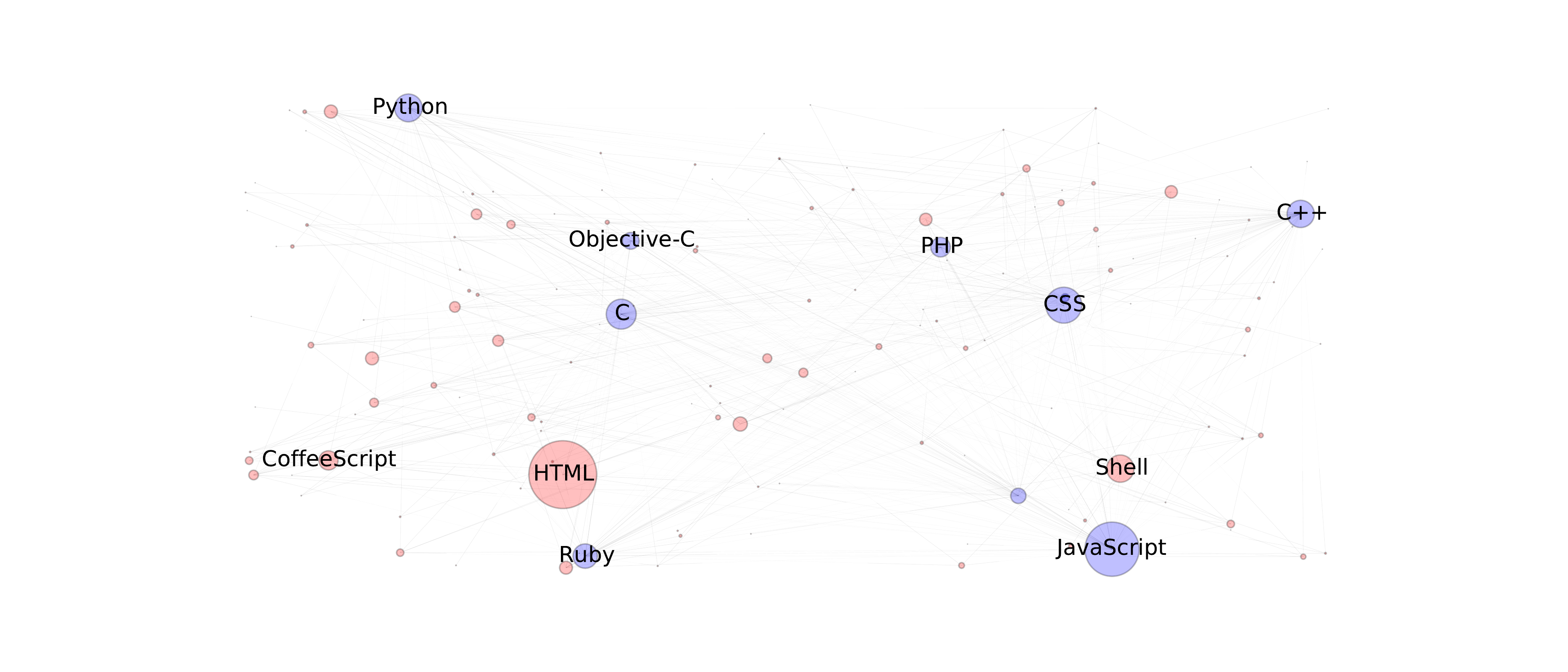}
\caption{Weighted directed graph of programming languages relationship in the top 1000 repositories. The blue nodes are the top 10 programming language from Table~\ref{t:pl-rank} and the red nodes are the other auxiliary programing languages used in the top 1000 repositories.}
\label{f:top100}
\end{figure*}

As we can see in Table~\ref{t:top100-10pgrank}, which is based on the directed graph described above, HTML is used in many projects and is used with other programming languages, followed by scripting languages to facilitate tasks in the projects. For example, Makefile scripts are a common practice in Unix based Operating Systems software developments to perform tedious tasks such as compilation and installation of the software from source code. Table~\ref{t:top100-10pl}, summarizes the top 10 programming languages that are used as auxiliary programming language in repositories. It is based on the number of projects, that the auxiliary languages are used in. Shell scripting is used in 47\%, and HTML is used in 42\% of the projects. To further investigate the dynamics of programming languages we build the weighted graph of the programming languages by considering the percentage of the code that is written in an auxiliary programming language. For example, if a project in written in Python (40\%), C (30\%) and Java (20\%), then the edge from Python would have the weight $w_{python\rightarrow java} = w_{python\rightarrow java} + 0.2$, and the edge from Python to C would have the weight, $w_{python\rightarrow c} = w_{python\rightarrow c} + 0.3$. Figure~\ref{f:top100}, depicts the weighted graph of programming languages, where the blue nodes are the top 10 programming language from Table~\ref{t:pl-rank} and the red nodes are the auxiliary programing languages used in the top 1000 repositories but are not in the Table~\ref{t:pl-rank}. As we can see, HTML, CSS, and Java Script are the dominant auxiliary programming languages based on the percentage of lines of code, followed by scripting languages such as Shell and general purpose and scripting languages such as Python.

\begin{table}[t]
  \centering
  \begin{tabular}{| c | c |}
    \hline
     Programming Language & Percentage of the Projects  \\ \hline
    Shell		&	47.1		\\ \hline
    HTML		&	41.6		\\ \hline
	JavaScript	&	39.2		\\ \hline
	CSS			&	38.1		\\ \hline
	Ruby 		&	29.8		\\ \hline
	Python 		&	28.8		\\ \hline
	C 			&	26.5		\\ \hline
	Makefile	&	25.5		\\ \hline
	C++			&	22.7		\\ \hline
	Objective-C	& 	19.2		\\ \hline
  \end{tabular}		
  \caption{Top 10 Programming Language, based on the number of projects they are used. As we can see, large percentage of the repositories rely on Shell scripts and use HTML.}
  \label{t:top100-10pl}
\end{table}
\section{Related Work}\label{s:related}


Prior work have studied and collected data from GitHub and other source code hosting repositories~\cite{Ding:2012:HOS:2351676.2351735},~\cite{Pham:2013:CSU:2486788.2486804}. However, to the best of our knowledge we are the first to collect and analyze GitHub's data at its entirety and at this scale. This dataset enabled us to study unique aspects of GitHub ecosystem, that was not possible previously. For example, GHTorrent~\cite{Gousi13}, is a mirror of GitHub's public events, using its RESTful API. GitHubarchive~\cite{gharchive}, collects and mirrors GitHub's public events since December 2011. Ray et al.~\cite{Ray:2014:LSS:2635868.2635922}, study the impact of programming languages on software quality. They collected data from 729 repositories on GitHub.

Another direction of research is on trends of different programming languages. For example, Meyerovich and Rabkin ~\cite{meyerovich2013empirical}, investigate the adoption of different programming languages, by using surveys, and collecting the metadata of some project from SourceForge and Ohloh. They find out that languages adoption follows power law, and factors such as libraries and existing code are more important to the developers, compared to performance and semantics. Chen et al.~\cite{1438333}, look at the software engineering and programming languages trends. They choose 17 and measure their evolution using different factors such as intrinsic, extrinsic and quantifying factors. Karus and Gall~\cite{Karus:2011:SLU:1985441.1985447} look at the language evolution of open source software development and the amount of code written in different languages. They study 22 open source softwares, and look at how XML and XSL are used. Another study explores the developer commit patterns in GitHub~\cite{6754372}, by defining four metrics to measure commit activity and code evolution: the changes in commits; the time between two commits; the author of each change; and the source code dependency.

Previous works look at the social aspects of coding and platform such as GitHub~\cite{yu2014exploring},~\cite{padhye2014study}. For example, Begel et al.~\cite{Begel:2013:SNM:2478553.2478856}, look at the social aspects of software development in platforms such as GitHub and MSDN, by interviewing leaders of companies. Marlow et al.~\cite{Marlow:2013:IFO:2441776.2441792}, look at impression formation, by tracing activity and personal profiles in GitHub. Authors conclude that developers form impressions around one another based on history of one's contributions across projects and interactions in the community.  Thung et al.~\cite{6498480} look at the network structure of social coding in GitHubg by constructing the developer-developer and project-project relationship graphs to examine characteristics of the graphs. The authors collected data from 100000 projects, and 30000 developers in this study. Vasilescu et al.~\cite{vasilescu2014continuous} look at the continuous integration in GitHub ecosystem, and explore whether direct and indirect continuations and different project characteristics such as the project age are associated with the success of the automatic builds.

\section{Future Work}\label{s:future}

In this section, we look at the future work based on the collected dataset. For example, investigation questions such as, ``What makes a programming language popular?'', ``How are programming languages adopted in projects?'', and ``What are the important internal and external factors in the adoption of a programming language?''

Another interesting research question is the investigation and discovery of users' hidden social structures based on their contributions to repositories. Such study requires the examination of the user-user and user-repository relationships. Furthermore, our goal is to study the users' expertise based on their contributions to different categories of applications. Imagine, a community of users who are only contributing to the projects related to data science, and another community that is only active in the projects related to security. Discovering the existence of such cliques, and investigating them is an interdisciplinary study, which uses social sciences and software engineering principles. Another direction of research is the examination of the dynamics of a project in its life span. For instance the study of the key moments and the tipping point in the success and adoption of a project between users.

\section{Conclusion}\label{s:conclusion}

In this work, we collected and studied GitHub, in its entirety. We investigated around 10 million user profiles, over 16 million public repositories, 11 million user relationships, and over 3 billion contributions. To the best of our knowledge, we are the first to perform a study at such a large scale. We looked at the growth and adoption rate of GitHub, based on the number of users and repositories. We provided an analysis of the top repositories, and the use of different programming languages for different purposes in projects. Additionally, we identified the popular programming languages on GitHub, and compared the results with other sources, such as TIOBE and Stack Overflow. Next, we investigated the correlation between different programming languages, and clustered the different languages and technologies using machine learning, and unsupervised learning. We found out that two clear clusters of programming languages separate from the remaining, ``web programming'' languages (Java Script, Ruby, PHP, CSS), and a second for ``system oriented programming''. We provided a hierarchical user-driven phylogenetic tree classification of programming languages. Furthermore, we studied the top 1000 repositories in more detail, by looking at the main and the auxiliary programming languages in these repositories. Our results indicated the use of multiple programming language in modern software artifacts. We provided a ranking of these auxiliary programming languages using different metrics, such as percentage of lines of code, and PageRank. We hope this work, adds to the body of literature on mining software repositories, and to answer research questions and shed new light on the dynamics of software development and programming languages.


\bibliographystyle{abbrv}
\bibliography{bibliography}

\end{document}